\documentclass[11pt]{article}
\usepackage[a4paper,margin=28mm]{geometry}
\usepackage[T1]{fontenc}
\usepackage{lmodern}
\usepackage{amsmath,amssymb,amsthm,mathtools}
\usepackage{microtype,xcolor}
\usepackage{pgfplots}
\pgfplotsset{compat=1.18}
\usepackage[colorlinks=true,linkcolor=blue!50!black,citecolor=blue!50!black,urlcolor=blue!50!black]{hyperref}
\hypersetup{pdftitle={Small-Space Streaming Hamming Approximation},
  pdfauthor={Przemysław Uznański}}

\newtheorem{theorem}{Theorem}[section]
\newtheorem{lemma}[theorem]{Lemma}
\newtheorem{proposition}[theorem]{Proposition}
\newtheorem{corollary}[theorem]{Corollary}
\newcommand{\E}{\mathbb E}
\newcommand{\Var}{\operatorname{Var}}
\newcommand{\Ham}{\operatorname{Ham}}
\newcommand{\softO}{\widetilde O}
\newcommand{\eps}{\varepsilon}
\newcommand{\ind}[1]{\mathbf 1_{\{#1\}}}
\newcommand{\ceil}[1]{\left\lceil #1\right\rceil}
\newcommand{\floor}[1]{\left\lfloor #1\right\rfloor}

\title{Small-Space Streaming Hamming Approximation}
\author{Przemys\l{}aw Uzna\'{n}ski\\[2pt]\small Pathway}
\date{17 November 2026}

\begin{document}
\maketitle

\begin{abstract}
For relative-$\eps$ Hamming matching with a preprocessed length-$m$
pattern and a streaming text, we give space bounds
\[
 \eps^{-8/5}m^{2/5+o(1)},\qquad
 \eps^{-1}m^{1/2+o(1)}\quad\text{words}.
\]
All retained pattern information counts toward space, and answers
are immediate and simultaneously correct with high probability.
The prior space bounds are $\softO(\eps^{-3/2}\sqrt m)$ and
$\softO((m/\eps)^{2/3})$, the latter an unpublished consequence of earlier
work~\cite{chan-golan-kociumaka-kopelowitz-porat-2020-approximating-hamming,clifford-kociumaka-porat-2018-streaming-k-mismatch}.
For $\eps=\Theta(m^{-a})$, our better bound improves on both leading
space exponents for every fixed $0<a<1/2$.

A Fourier matcher uses $\softO(\eps^{-2}(m/K)^2)$ words to estimate
each distance $d$ within $\eps(d+K)$, giving relative error at most
$2\eps$ for $d\ge K$. A threshold matcher based on random-prime
residue refinements and multi-stream dictionary
matching~\cite{golan-kopelowitz-porat-2018-multistream-dictionary}
gives the second bound directly and the first when combined with
the Fourier matcher. Both space bounds extend to $L_1$ matching over polynomially bounded
integer symbols.
\end{abstract}

\section{Introduction and results}
\label{sec:introduction}

Given a pattern $P\in\Sigma^m$ and a text $T\in\Sigma^n$, the
text-to-pattern Hamming distances are
\[
 d_s=\Ham(P,T[s:s+m])
     =|\{j\in[0,m):P[j]\ne T[s+j]\}|,
 \qquad 0\le s\le n-m.
\]
Indices are zero-based, and $T[a:b]=T[a]\cdots T[b-1]$ denotes a
half-open substring. Intervals used as index sets contain only integers.
In the streaming
setting, the answer for $s$ must be reported when $T[s+m-1]$ arrives.
A relative-$\eps$ estimate $\widehat d_s$ satisfies
$|\widehat d_s-d_s|\le\eps d_s$; in particular, zero distances must
be reported exactly.

\paragraph{Model.}
The pattern may be preprocessed in polynomial time and workspace before
the text starts. During text processing, every retained representation
of the pattern counts toward space, and the original pattern is no
longer available. The text arrives once and is fixed independently of
the algorithm's randomness. We assume $m\ge2$, $n\ge m$, and that
$n$ and $|\Sigma|$ are polynomial in $m$, with
$\Sigma=\{0,\ldots,|\Sigma|-1\}$. We use $O(\log(mn))$-bit words.
All answers must be correct simultaneously with probability at least
$1-n^{-c}$ for any fixed $c>0$; this is what we mean by
\emph{with high probability}.
Space and worst-case arrival time refer to the text-streaming phase.
We allow $m^{-C}\le\eps\le1/2$ for any fixed $C>0$.
The notation $\softO$ hides polylogarithmic factors in $m$, $n$, and
$1/\eps$.

Our first ingredient estimates large distances using Fourier samples.

\begin{theorem}\label{prop:fast-fourier}
For every integer $1\le K\le m$, the above model admits a streaming
estimator using $\softO(\eps^{-2}(m/K)^2)$ words and
$\softO(\eps^{-1}+\eps^{-3}m/K^2)$ worst-case time per arrival.
Its estimates satisfy $|E_s-d_s|\le\eps(d_s+K)$ simultaneously
at all alignments with high probability.
\end{theorem}

For $d_s\ge K$, this gives relative error at most $2\eps$.
Our second ingredient uses a shared representation of pattern residues
and text residue streams to estimate distances up to a chosen threshold.

\begin{theorem}\label{thm:shared}\label{lem:truncated-threshold}
For every integer $1\le H\le m$, there is a streaming Hamming matcher
using $\eps^{-1}\sqrt{\min\{m,H/\eps\}}\,m^{o(1)}$ words and
$\eps^{-2}m^{o(1)}$ worst-case time per arrival. With high probability,
it gives relative-$\eps$ estimates simultaneously at all alignments
with $d_s\le H$. In particular, $H=m$ gives relative-$\eps$ estimates
at all alignments using $\eps^{-1}m^{1/2+o(1)}$ words.
\end{theorem}

Combining this threshold matcher with the Fourier matcher gives a
second space bound for all alignments.

\begin{theorem}\label{thm:truncated}
Relative-$\eps$ Hamming matching in our model admits space
$\eps^{-8/5}m^{2/5+o(1)}$ and worst-case time
$\eps^{-2}m^{o(1)}$ per arriving text symbol.
\end{theorem}

Choosing between the two relative-approximation algorithms and explicit storage of the pattern
and current window gives space
\begin{equation}
 \softO\!\left(\min\left\{
   \eps^{-8/5}m^{2/5+o(1)},\
   \eps^{-1}m^{1/2+o(1)},\ m
 \right\}\right).                              \label{eq:combined-space}
\end{equation}
The $o(1)$ term depends only on $m$ and tends to zero as $m\to\infty$.
Balancing the Fourier matcher with the small-distance matcher of
Chan et al.~\cite[Theorem~B.5]{chan-golan-kociumaka-kopelowitz-porat-2020-approximating-hamming}
gives $\softO(m^{2/5}\eps^{-2})$ space, as shown in Section~\ref{sec:truncated}.

\paragraph{Prior bounds.}
Starikovskaya, Svagerka, and Uzna\'{n}ski~\cite{starikovskaya-svagerka-uznanski-2020-lp-stream}
give an algorithm using $\softO(\eps^{-2}\sqrt m)$ space and
$\softO(\eps^{-2})$ arrival time in the stronger model where the
pattern also arrives in one pass.
Chan, Golan, Kociumaka, Kopelowitz, and
Porat~\cite{chan-golan-kociumaka-kopelowitz-porat-2020-approximating-hamming}
use $\softO(\eps^{-3/2}\sqrt m)$ space and $\softO(\eps^{-3})$
arrival time with pattern preprocessing.
An unpublished consequence of prior work gives $\softO((m/\eps)^{2/3})$
space.\footnote{\label{fn:prior-space}Set $K=\ceil{(m/\eps)^{2/3}}$.
If $K\ge m/4$, explicit storage suffices. Otherwise run the exact
$K$-mismatch matcher of Clifford, Kociumaka, and Porat~\cite[Theorem~1.2]{clifford-kociumaka-porat-2018-streaming-k-mismatch}
and the fixed-scale matcher of Chan et al.~\cite[Section~2 and Theorem~B.2]{chan-golan-kociumaka-kopelowitz-porat-2020-approximating-hamming}
at scales $k=2^jK\le m$, $j\ge0$, with accuracy $\eta=\min\{\eps,1/3\}$.
Return the exact answer up to $K$. For $d_s>K$, return any estimate in
its scale's interval $[(1-\eta)k,2(1+\eta)k]$. On simultaneous success
every such estimate has relative error at most $\eta$, and the scale
with $k\le d_s<2k$ supplies one. The exact matcher uses $\softO(K)$ words,
and summing $\softO(m/(\eta\sqrt{k}))$ over the scales gives
$\softO(m/(\eps\sqrt K))=\softO(K)$ more.}

Our Fourier matcher recovers the same bound by balancing its
$\softO(\eps^{-2}(m/K)^2)$ space with the exact matcher's $\softO(K)$
at $K=\ceil{(m/\eps)^{2/3}}$. When $K<m/4$, use the exact answer
when available and Theorem~\ref{prop:fast-fourier} with parameter $\eps/2$
otherwise, since then $d_s>K$. For $K\ge m/4$, explicit storage suffices.

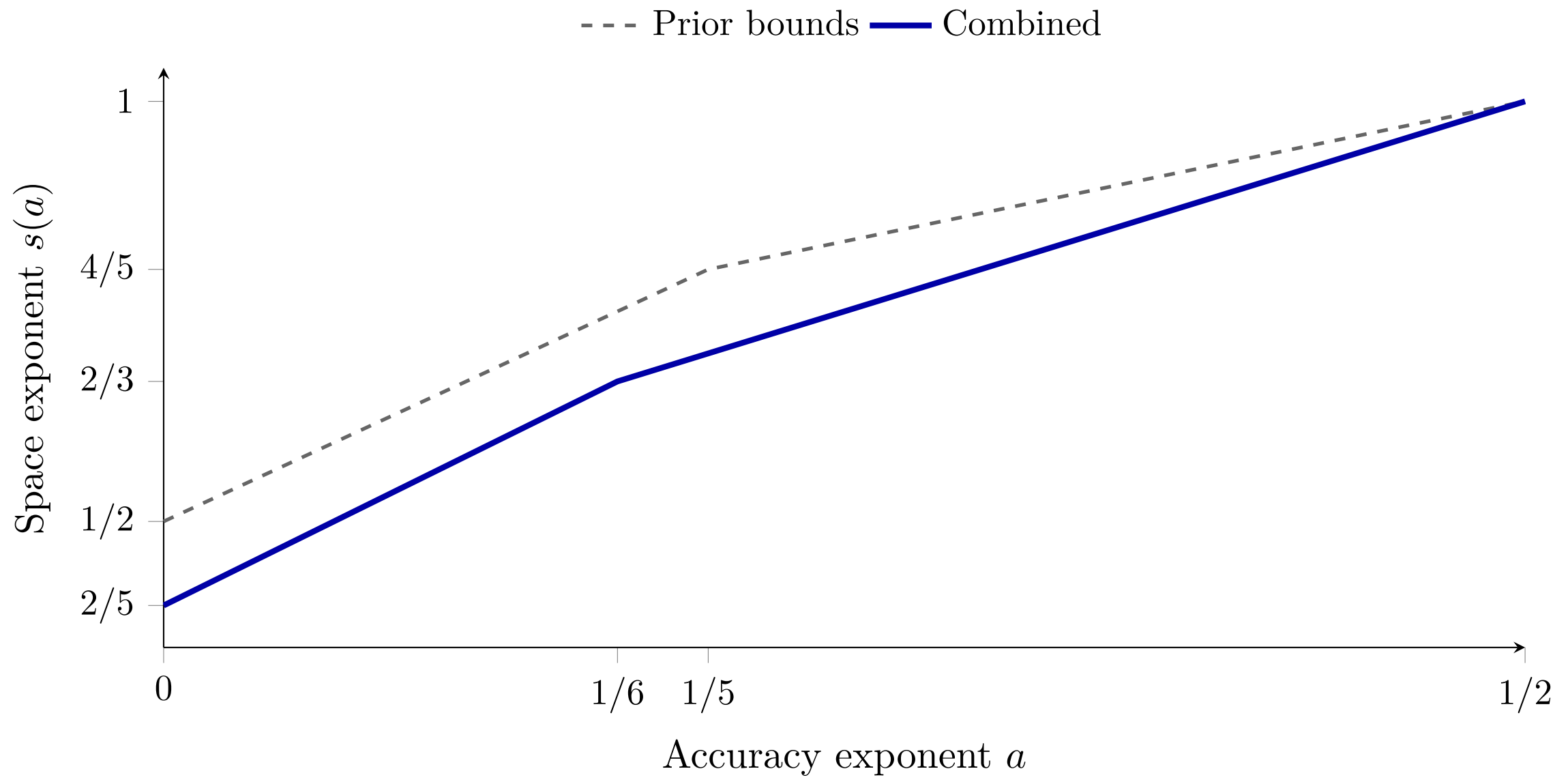
\begin{figure}[htbp]
\centering
\begin{tikzpicture}
\begin{axis}[
  width=.96\linewidth, height=7.2cm,
  xmin=0, xmax=.5, ymin=.35, ymax=1.04,
  axis lines=left,
  xlabel={Accuracy exponent $a$},
  ylabel={Space exponent $s(a)$},
  xtick={0,.1666666667,.2,.5},
  xticklabels={$0$,$1/6$,$1/5$,$1/2$},
  ytick={.4,.5,.6666666667,.8,1},
  yticklabels={$2/5$,$1/2$,$2/3$,$4/5$,$1$},
  tick label style={font=\small},
  label style={font=\normalsize},
  legend columns=2,
  legend style={at={(.5,1.02)},anchor=south,draw=none,
                fill=none,font=\small,cells={anchor=west}},
  tick align=outside,
]
\addplot[gray!80!black,dashed,line width=1pt,mark=none]
  coordinates {(0,.5) (.2,.8) (.5,1)};
\addlegendentry{Prior bounds}
\addplot[blue!65!black,line width=1.6pt,mark=none]
  coordinates {(0,.4) (.1666666667,.6666666667) (.5,1)};
\addlegendentry{Combined}
\end{axis}
\end{tikzpicture}
\caption{Space $m^{s(a)+o(1)}$ for $\eps=\Theta(m^{-a})$.
The dashed curve follows the published bound of Chan et al.\
up to $a=1/5$ and the unpublished bound in footnote~\ref{fn:prior-space} thereafter.
The solid curve gives~\eqref{eq:combined-space}, switching at $a=1/6$.}
\label{fig:space-comparison}
\end{figure}

Figure~\ref{fig:space-comparison} compares~\eqref{eq:combined-space}
with the minimum of these prior bounds and linear storage.
For $\eps=\Theta(m^{-a})$, our leading space exponent is smaller for
every fixed $0<a<1/2$.

For Theorem~\ref{prop:fast-fourier}, we first analyze a simpler implementation.
At an integer cutoff $1\le K\le m$, it uses
$\softO(\eps^{-2}(m/K)^2)$ space and
$\softO(\eps^{-2}m/K)$ arrival time, with mixed relative and additive
error $O(\eps(d_s+K))$. It stores Fourier sums for text blocks of
length $K$; zero padding makes the extra characters outside the window
contribute only to the variance.

The residue threshold matcher of Theorem~\ref{thm:shared} multiplies
the modulus by a random prime at each step and samples the increases in the number of occupied mismatch
residues. Pair collisions control both the sampling variance and the
deficit when refinement stops early. Exact dictionary matching evaluates
the sampled increases. One affine-grid sampler, with a common sample
budget across levels, lists the currently aligned pairs in bounded time.

Fourier-energy sampling has also been used in streaming sketches for
periodicity~\cite[Section~3.3]{crouch-mcgregor-2011-periodicity-linear-sketches}.
Here the centered moment bound and the causal block representation
provide the needed text-to-pattern estimator.

Section~\ref{sec:estimator} proves the Fourier matcher with the same
space and a simpler, slower implementation.
Section~\ref{sec:shared} proves Theorem~\ref{thm:shared}.
Section~\ref{sec:truncated} combines the two matchers to prove
Theorem~\ref{thm:truncated}.
Appendix~\ref{sec:faster} proves the static
sampling tool and the faster Fourier implementation of
Theorem~\ref{prop:fast-fourier}.
Appendices~\ref{sec:precision} and~\ref{sec:l1}
give the Fourier precision argument and the $L_1$ reduction.

\section{The Fourier matcher}
\label{sec:estimator}

For a vector $x\in\mathbb R^N$, write $\omega_N=\exp(2\pi\mathrm i/N)$
and $\widehat x_k=\sum_{j=0}^{N-1}x_j\omega_N^{-kj}$ for its unnormalized
Fourier transform. We estimate $m-\langle x,w\rangle$ by sampling a
frequency. Frequencies that carry more of the pattern energy must be
sampled with sufficiently high probability.

\begin{lemma}\label{lem:fourier}
Let $x,w\in\mathbb R^N$ with $\|x\|_2^2=m$, and put
$a=\widehat x$, $b=\widehat w$. Suppose $p$ is a distribution with
$p_k>0$ and $|a_k|^2/(Np_k)\le4m$ for every $k$. Draw $k$ from $p$
and put
\[
 Y=\frac{\operatorname{Re}(a_k\overline{(a_k-b_k)})}{Np_k}.
\]
Then $\E Y=m-\langle x,w\rangle$ and $\E Y^2\le4m\|x-w\|_2^2$.
\end{lemma}
\begin{proof}
In the mean, $p_k$ cancels the denominator, leaving
$N^{-1}\operatorname{Re}\sum_k a_k\overline{(a_k-b_k)}
=m-\langle x,w\rangle$ by the Parseval identity. For the second moment,
\[
 \E Y^2\le\frac1N\sum_k\frac{|a_k|^2}{Np_k}|a_k-b_k|^2
 \le\frac{4m}N\sum_k|a_k-b_k|^2=4m\|x-w\|_2^2 .
\]
\end{proof}

Use a map $h:\Sigma\to[0,G)$ and represent each symbol $a$ by the
standard basis vector $f(a)=e_{h(a)}\in\mathbb R^G$. Extend $f$ to
strings by concatenation, denoted by $\circ$, and write $0^b$ for
$b$ numerical zeros. Let $W$ be the current window, let $V$ be the
$\ell\le m$ characters retained immediately before it, and suppose the
$r$ most recent characters of $W$ are omitted. At length $N=2mG$, put
\[
 x=f(P)\circ0^{mG},\qquad
 w=f(W[0:m-r])\circ0^{rG}\circ0^{(m-\ell)G}\circ f(V).
\]
Let $D$ count mismatches under $h$ in the retained prefix. A matching
pair contributes one to the inner product, a mismatching pair contributes
two to the squared difference, and each unpaired symbol contributes one
to the squared difference. Thus
$\langle x,w\rangle=m-r-D$ and $\|x-w\|_2^2=2D+\ell+r$.
Lemma~\ref{lem:fourier} gives
\begin{equation}
 \E(Y-r)=D,\qquad \Var(Y-r)\le4m(2D+\ell+r). \label{eq:alphabet-base}
\end{equation}
Appendix~\ref{sec:precision} constructs an exactly sampleable dyadic
distribution $p$ with the required bound from the pattern alone.

\begin{proposition}\label{prop:coarse}
For every integer $1\le K\le m$ and accuracy $\eps$ in the model of
Section~\ref{sec:introduction}, there is a streaming estimator using
$\softO(\eps^{-2}(m/K)^2)$ words and
$\softO(\eps^{-2}m/K)$ worst-case time per arrival that reports
estimates $E_s$ satisfying
\begin{equation}
 |E_s-d_s|\le\eps(d_s+K)                         \label{eq:mixed}
\end{equation}
simultaneously for all alignments $s$, with high probability.
\end{proposition}
\begin{proof}
Take $G=|\Sigma|$ and $h(a)=a$, so the one-hot encoding preserves
every mismatch. Partition the text into blocks of length $K$. When
$t=s+m-1$ arrives, the blocks meeting the window also cover the
preceding string $V=T[s_0:s]$, where $s_0=K\floor{s/K}$. Thus $\ell=s-s_0<K$,
$r=0$, and $D=d_s$. By~\eqref{eq:alphabet-base}, the mean of
$L=\ceil{48m/(\eps^2K)}$ independent samples is unbiased and has variance
\[
 \frac{\Var Y}{L}\le\frac{\eps^2}{12}K(2d_s+K)
                     \le\frac{\eps^2}{12}(d_s+K)^2.
\]
Chebyshev's inequality bounds the error by $\eps(d_s+K)/2$ with failure
probability at most $1/3$. Take the median of $O(\log n)$ independent
means and union bound over alignments. Numerical error contributes at
most $\eps K/2$ by Appendix~\ref{sec:precision}, giving~\eqref{eq:mixed}.

During preprocessing, retain for each sampled frequency $k$ the two
coefficients $|a_k|^2/(Np_k)$ and $a_k/(Np_k)$ needed for $Y$.
Each symbol has just one nonzero coordinate. Consequently, even over
the full alphabet, the contribution of an arriving symbol to the
conjugated text transform at $k$ is a single phase. For each live
block $B$ keep
\begin{equation}
 Q_k(B,t)=\sum_{u\in B:\,u\le t}\omega_N^{k(Gu+T[u])}. \label{eq:block-sum}
\end{equation}
Maintain the sum over live blocks, starting with an empty queue.
Update the current block and the total on arrival. When $s_0$ advances,
subtract and discard the oldest block. Multiplying the total by
$\omega_N^{-Gks}$ gives $\overline{\widehat w_k}$, since a retained
symbol at $u$ occupies coordinate $(G(u-s)+T[u])\bmod N$ in $w$.
There are $O(m/K)$ live blocks per frequency and
$\softO(\eps^{-2}m/K)$ sampled frequencies. Scanning these frequencies
per arrival gives the stated time and space. Only pattern preprocessing
uses the full transform of length $2m|\Sigma|$.
\end{proof}

The pattern-dependent Fourier samples use offline preprocessing.
Obtaining the same matcher with one-pass pattern preprocessing
remains open in this approach.

\section{A residue threshold matcher}
\label{sec:shared}

To prove Theorem~\ref{thm:shared}, we use the following sampling lemma
from Appendix~\ref{sec:fast-residue}. Its resource bounds hold for every seed.

\begin{lemma}\label{lem:paired-sampling}
Let $Q$ be prime and let $M,t$ be integers with
$1\le M\le Q^2$ and $1\le t\le Q$.
There are random sets $A,B\subseteq[0,M)$, each of size at most $Qt$,
represented in $O(Qt)$ words, with the following properties.
For each integer shift $s$, the representation enumerates
$C(s)=\{z\in A:(s+z)\bmod M\in B\}$ in $O(t^2)$ time.
For fixed nonnegative weights $w_z$, put $q=(t/Q)^2$ and
$Y=q^{-1}\sum_{z\in C(s)}w_z$. Then
\[
 \E Y=\sum_z w_z,\qquad
 \Var Y\le\tfrac83(q^{-1}-1)\sum_z w_z^2.
\]
\end{lemma}

\begin{proof}[Proof of Theorem~\ref{thm:shared}]
We handle bounded $m$ by explicit storage. For sufficiently large $m$, put
\[
 R=2^{\ceil{\sqrt{\log_2 m\,\log_2\log_2 m}}},\qquad
 h=\ceil{\log_R m}.
\]
Let $\mathcal P$ be the primes in $[R,2R)$ and put
$\gamma=h/|\mathcal P|$. The standard prime-counting bound gives
$1/R\le\gamma=O(\log m/R)<1$.
Choose independent uniform $p_1,\ldots,p_h\in\mathcal P$ and set
$M_0=1$ and $M_i=p_1\cdots p_i$. Thus $M_h\ge m$.
For every fixed integer $0<x<m$,
\begin{equation}
 \Pr[M_i\mid x]\le\gamma^i.                    \label{eq:prime-prefix}
\end{equation}
Indeed, conditional on $M_{i-1}\mid x$, the remaining quotient
$x/M_{i-1}<m$ has at most $h$ prime divisors in $\mathcal P$,
since each is at least $R$ and $R^h\ge m$.
The next independent prime divides it with probability at most
$\gamma$. This also allows repeated primes.

Choose the depth
\[
 \ell=\min\left\{h,\ceil{\log_{1/\gamma}(6H/\eps)}\right\}.
\]
It depends only on $m,H,\eps$.
Fix an alignment $s$, its mismatch set $E\subseteq[0,m)$, and
$d=|E|\le H$. Let $H_i=|E\bmod M_i|$.
For $i<\ell$ and $b\in[0,M_i)$, let $g_{i,b}$ be one less than
the number of occupied residues modulo $M_{i+1}$ inside residue $b$
modulo $M_i$, or zero if it is empty. Then
\begin{equation}
 H_\ell=H_1+\sum_{i=1}^{\ell-1}\sum_b g_{i,b},
                                                        \label{eq:occupancy-telescope}
\end{equation}
since $\sum_b g_{i,b}=H_{i+1}-H_i$.
Each occupied child contains a mismatch, so $g_{i,b}^2$ is at most
the number of ordered pairs of distinct mismatches in the parent.
Similarly, $d-H_\ell$ is at most the number of unordered pairs
colliding modulo $M_\ell$. By~\eqref{eq:prime-prefix},
\[
 \E\sum_b g_{i,b}^2\le\gamma^i d(d-1),\qquad
 \E[d-H_\ell]\le\tfrac12\gamma^\ell d(d-1).
\]
If $\ell<h$, the second bound is at most $\eps d/12$.
If $\ell=h$, the deficit is zero because $M_h\ge m$.

\paragraph{The estimator.}
For $1\le i<h$, put $U_i=(2R)^i\ge M_i$ and fix a prime $Q_i$
with $Q_i^2\ge U_i$ and $Q_i^2=O(U_i)$, using Bertrand's postulate.
We use the common sample budget
\[
 S=1+\tfrac83\sum_{i=1}^{h-1}Q_i^2\gamma^i,\qquad
 t_\ast=\ceil{4\sqrt S/\eps}.
\]
Set $t_i=\min\{Q_i,t_\ast\}$ and $q_i=(t_i/Q_i)^2$.
Independently at each level $i<\ell$, draw $A_i,B_i$ from
Lemma~\ref{lem:paired-sampling} with parameters $M_i,Q_i,t_i$.
Set
\[
 Z=H_1+\sum_{i=1}^{\ell-1}q_i^{-1}
       \sum_{b\in A_i:\ (s+b)\bmod M_i\in B_i}g_{i,b}.
\]
Conditional on the moduli, its mean is $H_\ell$.
A fully sampled level has $q_i=1$; otherwise $q_i^{-1}=Q_i^2/t_\ast^2$.
The sampling lemma and independence between levels therefore give
\begin{align}
 \E[(Z-H_\ell)^2]
 &\le\tfrac83 d(d-1)\sum_{i<\ell}(q_i^{-1}-1)\gamma^i \notag\\
 &\le\frac{S d(d-1)}{t_\ast^2}
 \le\frac{\eps^2d(d-1)}{16}.                \label{eq:refinement-variance}
\end{align}
Consequently, Cauchy--Schwarz and the deficit bound give
\begin{equation}
 \E|Z-d|
 \le\sqrt{\E[(Z-H_\ell)^2]}+\E[d-H_\ell]
 \le\frac{\eps d}{4}+\frac{\eps d}{12}
 =\frac{\eps d}{3}.                         \label{eq:refinement-error}
\end{equation}
Markov's inequality gives relative-$\eps$ accuracy with probability
at least $2/3$ when $d>0$. At $d=0$ or $d=1$, the ideal answer is exact.

\paragraph{Streaming implementation.}
Use the equal-length multi-stream dictionary matcher of
Golan, Kopelowitz, and Porat~\cite[Theorem~1]{golan-kopelowitz-porat-2018-multistream-dictionary}.
For $D$ patterns and $v$ streams, it uses $\softO(D+v)$ words,
including the shared dictionary, and polylogarithmic worst-case time
per stream update. It reports the matching pattern identifier at the
current suffix, with high probability; duplicate patterns share an identifier.

At level $i$, for every $b\in A_i$ put all its nonempty fine pattern
residues into the dictionary: for $a=b+M_i j$, $0\le j<p_{i+1}$, the string is
\[
 P[a]\circ P[a+M_{i+1}]\circ P[a+2M_{i+1}]\circ\cdots,
\]
stopping before position $m$.
Their lengths are $\floor{m/M_{i+1}}$ or
$\ceil{m/M_{i+1}}$, so two equal-length dictionaries suffice.
For each $v\in B_i$, maintain its $p_{i+1}$ fine text residue streams,
with state for both dictionary lengths. An arrival at position $t$
updates stream $t\bmod M_{i+1}$ if $t\bmod M_i\in B_i$.
Cache each stream's latest matching identifier for each length.

At alignment $s$, enumerate the selected parent pairs using
Lemma~\ref{lem:paired-sampling}. For each reported parent $b$,
compare every nonempty fine pattern residue $a=b+M_i j$ against
the current suffix of stream $(s+a)\bmod M_{i+1}$, using its length.
This suffix consists exactly of the corresponding characters in
$T[s:s+m]$. The number of failed equalities is the number of
occupied children, and hence determines $g_{i,b}$.
Maintain $H_1$ in the same way, using all $M_1$ pattern and text
residues. All streams start at time zero, so every answer is immediate.
Sorted residue lists support all lookups in logarithmic time.

\paragraph{Resources and amplification.}
Each level retains at most $Q_i t_i$ pattern parents and as many
text parents, each with fewer than $2R$ children. Thus space is
$\softO(R+Rt_\ast\sum_{i<\ell}Q_i)$.
At most two dictionary states are updated per level per arrival.
The sampling lemma enumerates all active parents in $O(t_i^2)$ time;
their child queries and the root scan give worst-case arrival time
$\softO(R+Rh t_\ast^2)$.
Since $R\gamma=O(\log m)$ and $h=O(\sqrt{\log m/\log\log m})$,
\[
 S=O\!\left(1+\sum_{i<h}(2R\gamma)^i\right)
   =\exp(O(h\log\log m)).
\]
Thus $t_\ast=\eps^{-1}\exp(O(h\log\log m))$. For the space bound,
\[
 \sum_{i<\ell}Q_i=O\!\left(h(2R)^{(\ell-1)/2}\right)
 \le \gamma^{-(\ell-1)/2}\exp(O(h\log\log m)),
\]
using $2R\gamma=O(\log m)$. Substitution gives arrival time
$\eps^{-2}m^{o(1)}$ and space
\begin{equation}
 \softO\!\left(R+Rt_\ast\sum_{i<\ell}Q_i\right)
 \le\eps^{-1}\gamma^{-(\ell-1)/2}m^{o(1)}.    \label{eq:refinement-space}
\end{equation}
Both subpolynomial factors are at most $\exp(O(\sqrt{\log m\,\log\log m}))$,
since $\log R+h\log\log m=O(\sqrt{\log m\,\log\log m})$.
Finally, $\gamma\ge1/R$ and $R^{h-1}<m$ give the first bound below;
minimality of $\ell$ gives the second, including $\ell=1$ and $\ell=h$:
\[
 \gamma^{-(\ell-1)/2}<\sqrt m,\qquad
 \gamma^{-(\ell-1)/2}<\sqrt{6H/\eps}.
\]
This gives the stated space bound. When $H=m$, every $d_s$ is at most
$H$ and the bound becomes $\eps^{-1}m^{1/2+o(1)}$.

Every weight is either one or $Q_i^2/t_\ast^2$, so $t_\ast^2$ is a
common integer denominator. Since $(2R)^h=m^{1+o(1)}$ and
$t_\ast=\eps^{-1}m^{o(1)}$, all scaled weights and accumulated answers
are polynomially bounded integers and fit logarithmic words.
Preprocessing enumerates the candidate primes in polynomial time.
All uniform finite choices, including invertible matrices, can use
rejection sampling with constant acceptance probability. Capping each
draw at $O(\log n)$ trials adds an inverse-polynomial failure probability.
Every retained representation is charged, and the original pattern
is discarded before text streaming.

Take the median of $O(\log n)$ independent complete copies, including
independent primes and samples, and union bound over alignments.
The repetitions preserve both resource bounds. The moment calculation
assumes exact dictionary answers. Give dictionary failures a separate
inverse-polynomial budget, with seeds independent of the hierarchy
and sampling. On simultaneous dictionary success the implementation
agrees with the ideal process; its moments are not conditioned on this event.
\end{proof}

\section{Combining the residue and Fourier matchers}
\label{sec:truncated}

We prove Theorem~\ref{thm:truncated} by combining the residue matcher
at a smaller threshold with Fourier estimates for large distances.

\begin{proof}[Proof of Theorem~\ref{thm:truncated}]
Fix $1\le K\le m$. Run the Fourier matcher of
Theorem~\ref{prop:fast-fourier} at accuracy $\eps/3$ and
the residue matcher of Theorem~\ref{thm:shared} at threshold $H=\min\{m,2K\}$.
Output the Fourier estimate $E_s$ if $E_s\ge K$ and the threshold
answer otherwise. On simultaneous success, the Fourier guarantee gives
\[
 E_s\ge K\implies d_s\ge K/2,\qquad
 E_s<K\implies d_s<2K.
\]
Indeed, $d_s<K/2$ implies $E_s<K/2+\eps K/2<K$, whereas
$d_s\ge2K$ implies $E_s\ge2K-\eps K>K$.
In the first branch, $|E_s-d_s|\le\eps(d_s+K)/3\le\eps d_s$;
in the second, the threshold guarantee applies, including at zero.
The failure budgets add, as do all retained states and arrival costs.

The space is at most
\[
 \eps^{-3/2}\sqrt K\,m^{o(1)}
       +\softO\!\left(\eps^{-2}(m/K)^2\right).
\]
For $\eps\ge1/m$, choose $K=\ceil{m^{4/5}\eps^{-1/5}}\le m$.
Both terms are then bounded by $\eps^{-8/5}m^{2/5+o(1)}$.
Since
\[
 \frac{m}{\eps K^2}\le(m\eps)^{-3/5}\le1,
\]
Theorem~\ref{prop:fast-fourier} gives Fourier arrival time
$\softO(\eps^{-1}+\eps^{-3}m/K^2)=\softO(\eps^{-2})$; the threshold matcher
takes $\eps^{-2}m^{o(1)}$ time.
For $\eps<1/m$, explicit storage and direct comparison satisfy
both claimed bounds.
\end{proof}

\paragraph{The Fourier balance with an approximate threshold matcher.}
The same selection rule works with the threshold matcher of
Chan et al.~\cite[Theorem~B.5]{chan-golan-kociumaka-kopelowitz-porat-2020-approximating-hamming},
which uses $\softO(\eps^{-2}\sqrt K)$ words. At accuracy
$\eta=\min\{\eps,1/3\}$ its guarantee at scale $k'=d_s$ gives relative
accuracy for $1\le d_s\le H$, and its guarantee at $k'=1$ identifies
zero exactly by an answer below $1-\eta$. Report zero in that case.
Taking $K=\ceil{m^{4/5}}$ balances this space with
$\softO(\eps^{-2}(m/K)^2)$ from Proposition~\ref{prop:coarse}, giving
$\softO(m^{2/5}\eps^{-2})$ space.

\section*{AI disclosure}

The results of this paper were obtained in cooperation with GPT 6.1.
Model assisted in preparing this write-up. I take full responsibility
for the contents of this paper.

\newcommand{\etalchar}[1]{$^{#1}$}

\appendix
\section{Sampling and faster Fourier matching}
\label{sec:faster}

We prove the faster Fourier implementation of
Theorem~\ref{prop:fast-fourier}, then prove the
static sampling tool from Section~\ref{sec:shared}.
Following the block scheduling used by
Starikovskaya, Svagerka, and Uzna\'{n}ski~\cite[Section~4]{starikovskaya-svagerka-uznanski-2020-lp-stream},
we prepare contributions from older text before they are needed.
We omit a short suffix of recent text, charging its contribution to
the additive error allowance. Hashing to a small alphabet and using
one-hot encoding lets us share one alphabet map among all Fourier
samples in a repetition and batch both updates and queries.

\subsection{Short Fourier evaluations}
\label{sec:fast-tools}

We use short Fourier evaluations at nonuniform frequencies. The following
construction uses the Taylor expansion around an FFT grid of
Anderson and Dahleh~\cite[Section~3.1]{anderson-dahleh-1996-rapid-dft}.
We include the argument for both the evaluation map and its transpose.

\begin{lemma}\label{lem:short-fourier}
Let $B,L$ be positive integers and $z_1,\ldots,z_L$ be $N_0$th roots of
unity specified by their exponents. Suppose $B,L,N_0$
are polynomially bounded in $m$. For polynomially bounded complex
inputs, either of the maps
\[
 F_\ell=\sum_{v=0}^{B-1}x_vz_\ell^v\quad(1\le\ell\le L),
 \qquad
 H_v=\sum_{\ell=1}^Lc_\ell z_\ell^v\quad(0\le v<B)
\]
can be evaluated to any prescribed inverse-polynomial absolute error
in $\softO(B+L)$ time and space, using $O(\log(mn))$-bit words.
\end{lemma}
\begin{proof}
Choose a power of two $N$ with $4B\le N<8B$. Write
$z_\ell=\exp(\mathrm i\theta_\ell)$ and choose the nearest grid angle:
\[
 \theta_\ell=\frac{2\pi j_\ell}{N}+\delta_\ell\pmod{2\pi},
 \qquad |\delta_\ell|\le\frac{\pi}{N},
 \qquad \beta_\ell=B\delta_\ell.
\]
The indices $j_\ell$ are computed by integer arithmetic on the
exponents. Since $|\beta_\ell|\le\pi/4$, for $0\le v<B$ we have
\[
 z_\ell^v=e^{2\pi\mathrm i j_\ell v/N}
 \left(\sum_{r=0}^{p-1}
   \frac{(\mathrm i\beta_\ell)^r}{r!}(v/B)^r+\rho_{\ell,v}\right),
 \qquad
 |\rho_{\ell,v}|\le e^{\pi/4}\frac{(\pi/4)^p}{p!}.
\]
For $F$, compute the positive-exponent $N$-point Fourier sums of
$x_v(v/B)^r$, padded with zeros, and read coordinate $j_\ell$.
Multiply it by $(\mathrm i\beta_\ell)^r/r!$ and sum over $r$.
Transposing gives $H$: coordinate reads become bucket additions,
diagonal scalings are unchanged, and the Fourier matrix is symmetric.
Each Taylor term costs
$O(N\log N+L)$ operations.

The truncation error is bounded by the displayed remainder times
$\sum_v|x_v|$ or $\sum_\ell|c_\ell|$. Thus $p=O(\log(mn))$ suffices.
Compute Taylor weights recursively; their absolute sum is bounded,
and the normalized powers are at most one. Each FFT has logarithmic
depth and polynomial error amplification. Consequently
inverse-polynomial rounding precision suffices throughout.
Repeated nodes are allowed, and no matrix inversion is used.
\end{proof}

\subsection{Batched Fourier matching}
\label{sec:fast-implementation}

\begin{proof}[Proof of Theorem~\ref{prop:fast-fourier}]
If $\eps K<8$, Proposition~\ref{prop:coarse} suffices, since
$\eps^{-2}m/K<8\eps^{-3}m/K^2$.
Otherwise let $G$ be the smallest power of two at least $12/\eps$, and put
\[
 L=\ceil{96m/(\eps^2K)},\qquad
 B=\min\{\floor{L/G},\floor{\eps K/8}\}.
\]
Since $G<24/\eps$, we have $L/G>m/(\eps K)\ge2$, so $B\ge1$.

\paragraph{Omitting recent input.}
Partition time into working blocks $I_j=[jB,(j+1)B)$ and storage
blocks of length $C$, where
\[
 C=B\floor{K/(2B)},\qquad \delta=(1-m)\bmod B.
\]
For an output at $t\in I_j$, $t\ge m-1$, put $s=t-m+1$,
$v=(j-1)B$, and
\[
 s_0(t)=\max\{0,C\floor{(s-\delta)/C}\},\qquad
 \ell=s-s_0(t),\qquad r=t-v+1.
\]
We have $K/4\le C\le K/2$ and $B\le K/16$. Since both $C$ and
$m-1+\delta$ are multiples of $B$, each input block belongs to one
storage block and $s_0(t)$ is constant throughout each output block.
Moreover,
\[
 0\le\ell<C+B,\qquad r\le2B\le\eps K/4,\qquad
 \ell+r<C+3B<K.
\]
Thus $s_0(t)\le s\le v$. We omit the $r$ most recent symbols;
this changes the Hamming distance by at most $r$.

\paragraph{One hash and one average.}
Choose a uniform pairwise-independent map $h:\Sigma\to[0,G)$.
Independent affine Boolean maps for its output bits give a
polylogarithmic seed and evaluation time.
Fix the alignment, write $d=d_s$, and let $d_h$ be its full hashed
distance. The small-alphabet argument of
Kopelowitz and Porat~\cite[Section~2]{kopelowitz-porat-2018-simple-hamming-approximation}
gives $\E[d-d_h]=d/G$. Hence Markov's inequality gives
\[
 \Pr[d-d_h>\eps d/2]\le1/6;
\]
when $d=0$, the loss is zero.
Use the packed vectors of Section~\ref{sec:estimator} with
$V=T[s_0(t):s]$, the omitted suffix of length $r$, and this map $h$.
Write $D$ for the retained-prefix hashed distance. Conditional on $h$,
let $U$ be the average of $L$ independent copies of $Y-r$.
By~\eqref{eq:alphabet-base}, $\E[U\mid h]=D$ and
\begin{equation}
 \Var(U\mid h)\le\frac{4m(2D+\ell+r)}L
 \le\frac{\eps^2K(2d+K)}{24}
 \le\frac{\eps^2(d+K)^2}{24}.             \label{eq:group-variance}
\end{equation}
Chebyshev's inequality gives $|U-D|\le\eps(d+K)/2$ except with
probability $1/6$. Also $0\le d-D\le d-d_h+r$.
Reserving $\eps K/4$ for numerical error, both good events give
\[
 |\text{computed }U-d|
 \le\eps(d+K)/2+\eps d/2+r+\eps K/4\le\eps(d+K).
\]
Their joint failure probability is at most $1/3$. A median of
$O(\log n)$ independent copies, each with a fresh hash and frequency
list, gives the guarantee simultaneously at every alignment with
high probability.

\paragraph{Preparing output tables.}
For one copy with frequencies $k_1,\ldots,k_L$, put $N=2mG$,
$\eta=\omega_N$, and precompute
\[
 A=\frac1L\sum_{a=1}^L\frac{|\widehat x_{k_a}|^2}{Np_{k_a}},
 \qquad c_a=\frac{\widehat x_{k_a}\eta^{Gk_a(m-1)}}{LNp_{k_a}},
 \qquad z_a=\eta^{-Gk_a}.
\]
For the retained text let $Q_a=\sum_{u=s_0(t)}^{v-1}\eta^{k_a(Gu+h(T[u]))}$.
Since $\overline{\widehat w_{k_a}}=\eta^{-Gk_as}Q_a$ and $s=t-m+1$,
the average at time $t$ is
\begin{equation}
 U_t=A-r-\operatorname{Re}\sum_{a=1}^Lc_aQ_az_a^t.       \label{eq:group-table}
\end{equation}
The $Q_a$ are constant on $I_j$, and all their input is known at
the start of $I_{j-1}$. Store one moment vector per retained storage
block and their total. At that time the completed block $I_{j-2}$
is available. Expand it into $BG$ one-hot coordinates and apply
one forward transform at the nodes $\eta^{k_a}$, multiplied by its
global starting phase. Add these increments to its single storage
block and to the total. Delete the at most one block preceding the
new cutoff $s_0$. One transposed evaluation with nodes $z_a$ and
weights $c_aQ_az_a^{jB}$ then gives the entire table in~\eqref{eq:group-table}.

Since $BG\le L$, Lemma~\ref{lem:short-fourier} bounds the job's time and workspace by
$\softO(L+BG)=\softO(L)$. Only a working block is expanded during
streaming. Spread the job over the $B$ arrivals of $I_{j-1}$;
current outputs read the previously completed table. Cutoffs only
increase, so deleted moments are never needed again. The first two
blocks require no output because $m-1\ge2B$; initialize their
history as empty.

\paragraph{Resources and precision.}
There are $O(m/C+1)=O(m/K)$ retained storage blocks, each with $L$
moments. The hash seed, sampled coefficients, two raw buffers, two
output tables, and in-progress transforms use $\softO(L)$ further
words. Including all copies, space is
$\softO(Lm/K)=\softO(\eps^{-2}(m/K)^2)$ and worst-case arrival time is
\[
 \softO(1+L/B)=\softO(G+L/(\eps K))
             =\softO(\eps^{-1}+\eps^{-3}m/K^2).
\]
The spectra are computed during preprocessing and discarded after
sampling. Appendix~\ref{sec:precision} gives logarithmic-word arithmetic
with error at most $\eps K/4$ per copy, completing the error budget.
\end{proof}

\subsection{Sampling with an affine grid}
\label{sec:fast-residue}

\begin{proof}[Proof of Lemma~\ref{lem:paired-sampling}]
Identify $z\in[0,Q^2)$ with
$X(z)=(z\bmod Q,\floor{z/Q})^\top\in\mathbb F_Q^2$.
Choose a uniform invertible matrix $L\in\mathbb F_Q^{2\times2}$
and an independent uniform $c\in\mathbb F_Q^2$.
Put $F(X)=LX+c$ and $I=\{0,\ldots,t-1\}$, and set
\[
 A=\{z<M:F(X(z))_1\in I\},\qquad
 B=\{z<M:F(X(z))_2\in I\}.
\]
The inverse images of $I\times\mathbb F_Q$ and
$\mathbb F_Q\times I$ each have $Qt$ points. Enumerate these images,
restrict to $z<M$, and retain the lists, $L,c,L^{-1}$ in $O(Qt)$ words.

\paragraph{Moments.}
First draw the two rows of $L$ independently and uniformly from
$\mathbb F_Q^2$, before conditioning on invertibility.
For any fixed $L$, the two independent offsets give each $z$ paired
inclusion probability $q$, hence $\E[Y\mid L]=\mu:=\sum_z w_z$.
For distinct $z,w$, a uniform affine form takes independent uniform
values on $X(z),X(w)$ because their difference is uniform and the
offset makes the first value independently uniform. The second
form independently has the same property for the two distinct shifted
points. Thus the paired indicators are pairwise independent before
conditioning, giving $\E[(Y-\mu)^2]=(q^{-1}-1)\sum_z w_z^2$.
The probability of invertibility is
$(1-Q^{-2})(1-Q^{-1})\ge3/8$. Conditioning preserves the mean and
increases this second moment by at most $8/3$, proving the lemma's
variance bound.

\paragraph{Enumeration.}
Reduce $s$ modulo $M$. For $j\in\{0,1\}$ write
$s-jM=d_0+Qd_1$, with $0\le d_0<Q$.
Partition $[0,M)$ according to the wrap $j=\ind{z+s\ge M}$ and
carry $e=\ind{(z\bmod Q)+d_0\ge Q}$.
On region $(j,e)$, writing $z'=(z+s)\bmod M$, we have
\[
 X(z')=X(z)+\delta_{j,e}\quad\text{over }\mathbb F_Q,
 \qquad \delta_{j,e}=(d_0,d_1+e)^\top.
\]
Thus paired inclusion on this region is equivalent to
\[
 LX(z)+c_{j,e}\in I^2,\qquad
 c_{j,e}=c+(0,(L\delta_{j,e})_2)^\top.
\]
The regions depend only on $M,Q,s$.

For each region $(j,e)$ and each $(u,v)\in I^2$, compute
$X=L^{-1}((u,v)^\top-c_{j,e})$.
Using integer coordinates $0\le x,y<Q$, recover $z=x+Qy$ and
retain it exactly when $z<M$ and its actual wrap and carry equal
$(j,e)$. The region identity then proves that $z\in C(s)$.
Conversely, every active point has a unique region and pair $(u,v)$,
so it is reported exactly once. The $4t^2$ candidates take constant
time each using the stored inverse.
\end{proof}

\section{Logarithmic-word implementation}
\label{sec:precision}

For either Fourier implementation, let $a=\widehat x$ and $N=2mG$.
During preprocessing compute dyadic upper bounds
$|a_k|^2\le u_k\le |a_k|^2+1/2$ and put $v_k=1+\floor{u_k}$.
These bounds can be obtained by evaluating the DFT to sufficiently fine
inverse-polynomial precision, since $|a_k|\le m$.
The Parseval identity gives $\sum_k|a_k|^2=Nm$. Thus
\[
 v_k\ge\max\{1,|a_k|^2\},\qquad V=\sum_k v_k\le N(m+3/2).
\]
Let $M$ be the least power of two at least $V$. Add $M-V$ to the
first bin and divide the resulting masses by $M$ to obtain $p$.
Drawing $\log_2 M$ random bits selects a position in the cumulative
integer masses exactly. Since $M<2V\le4Nm$ for $m\ge2$, every bin has
$p_k\ge1/M$ and $|a_k|^2/(Np_k)\le M/N\le4m$, as required by
Lemma~\ref{lem:fourier}.

All stored sample coefficients are polynomially bounded, since
$1/p_k\le M$ and $N,m,1/\eps$ are polynomial in $m$.
Round these coefficients and the phases to a common dyadic precision,
reducing phase exponents modulo $N$ before evaluation. Each block sum
receives only additions and subtractions, so errors over polynomially
many updates accumulate additively. In the batched algorithm,
Lemma~\ref{lem:short-fourier} gives inverse-polynomial absolute error
in each forward transform. The output transform multiplies moment errors
by polynomially bounded coefficients, and its results are used only
for output. Thus a sufficiently fine inverse-polynomial mesh gives
absolute error at most $\eps K/2$ per coarse mean and $\eps K/4$ per
batched mean. Medians preserve these error bounds. Each scalar uses
$O(\log(mn))$ bits, and the full pattern spectrum and sampling tables
are discarded before the text arrives.

\section{Reduction from \texorpdfstring{$L_1$}{L1} distance}
\label{sec:l1}

Let $U\ge1$ be an integer. For symbols in $\{0,\ldots,U-1\}$, define
$D_s=\sum_{j=0}^{m-1}|P[j]-T[s+j]|$.
Exact reductions from $L_1$ to Hamming distance are given by
Labib, Uzna\'{n}ski, and Wolleb-Graf~\cite[Section~5]{labib-uznanski-wolleb-graf-2019}.
Signed reconstruction requires an additional stability argument for
relative approximation. We give a symmetric halving identity and use
a common random translation to bound its absolute coefficient mass.

\begin{corollary}\label{cor:l1}
In the model of Section~\ref{sec:introduction}, let $U=m^{O(1)}$,
$w=1+\ceil{\log_2 U}$, $J=w(w+1)$, and $\eta=\eps/(12J)$.
Suppose a streaming Hamming matcher on an alphabet of size $2^w$ returns
a relative-$\eta$ estimate with probability at least $2/3$ at each fixed
alignment. Then streaming $L_1$ matching on $\{0,\ldots,U-1\}$ gives
relative-$\eps$ estimates at all alignments simultaneously with high
probability, using $O(J\log J\log n)$ times the matcher's space and
worst-case arrival time at the same pattern and text lengths.
\end{corollary}

Since $J=O(\log^2(2U))$, applying the corollary to
Theorem~\ref{thm:truncated} and to Theorem~\ref{thm:shared} with $H=m$
gives $L_1$ space
$\eps^{-8/5}m^{2/5+o(1)}$ and $\eps^{-1}m^{1/2+o(1)}$, respectively,
with $\eps^{-2}m^{o(1)}$ worst-case arrival time in either case.
The Fourier balance in Section~\ref{sec:truncated} gives
$\softO(m^{2/5}\eps^{-2})$ space.

\begin{samepage}
\begin{lemma}\label{lem:l1-stability}
Let $U\ge2$ be an integer and put $w=1+\ceil{\log_2 U}$.
There are $J=w(w+1)$ maps
$f_\ell:\{0,\ldots,2^w-1\}\to\{0,\ldots,2^w-1\}$, computable
in constant word time, and
integer coefficients $c_\ell$ of magnitude $O(U)$ such that
\[
 |X-Y|=\sum_{\ell=1}^J c_\ell\ind{f_\ell(X)\ne f_\ell(Y)}
\]
for all integers $X,Y\in[0,2^w)$.
If $\Delta$ is uniform in $[0,2^{w-1})\cap\mathbb Z$, then, for every
fixed $x,y\in[0,U)\cap\mathbb Z$,
\begin{equation}
 \E_\Delta\sum_{\ell=1}^J|c_\ell|
    \ind{f_\ell(x+\Delta)\ne f_\ell(y+\Delta)}
 \le w(w+1)|x-y|.                              \label{eq:l1-mass}
\end{equation}
\end{lemma}
\end{samepage}

\begin{proof}
Write $q_i(z)=\floor{z/2^i}$, $b_i(z)=q_i(z)\bmod2$, and
$H_f=\ind{f(X)\ne f(Y)}$. Define two maps on bit pairs by
\[
 \begin{array}{c|rrrr}
 (a,b)&00&01&10&11\\\hline
 f_+(a,b)&0&2&2&1\\
 f_-(a,b)&2&0&1&2
 \end{array}
\]
Their Hamming distances satisfy
\[
 \ind{f_+(a,b)\ne f_+(a',b')}
 -\ind{f_-(a,b)\ne f_-(a',b')}
 =(a-a')(b-b').
\]
For $0\le i<k<w$, put
\[
 g_{i,k}^{\pm}(z)=3q_{k+1}(z)+f_\pm(b_k(z),b_i(z)).
\]
The required identity is
\begin{equation}
 |X-Y|=\sum_{i=0}^{w-1}2^i\left(
 H_{q_i}-H_{q_{i+1}}+
 \sum_{k=i+1}^{w-1}
       (H_{g_{i,k}^{+}}-H_{g_{i,k}^{-}})\right).  \label{eq:l1-identity}
\end{equation}
For $X\ne Y$, let $j$ be their highest differing bit and put
$\sigma=b_j(X)-b_j(Y)=\operatorname{sgn}(X-Y)$. The parenthesis at
bit $i$ equals $\sigma(b_i(X)-b_i(Y))$. Indeed, for $k<j$ the prefixes
above $k$ differ, so both maps disagree and their contributions cancel.
For $k>j$ the $k$th bits agree, so the bit-pair identity gives zero.
Thus if $i<j$, only $k=j$ contributes and gives the claimed product.
At $i=j$ the prefix-indicator difference is one, and for $i>j$ all
terms vanish. Summing with weights $2^i$ gives $\sigma(X-Y)=|X-Y|$.
For $X=Y$ all summands are zero.
There are $2w+2\binom w2=w(w+1)$ terms, each using the same map on
both inputs. Their values lie in $[0,2^w)$, since $k\ge1$ gives
$g_{i,k}^{\pm}(z)<3\cdot2^{w-2}<2^w$.

Keep these terms separately when taking absolute coefficients.
Every map at scale $i$ depends only on bits $i$ and above, and their
total absolute coefficient is $2^{i+1}(w-i)$. Thus the mass is at most
\[
 \sum_{i=0}^{w-1}2^{i+1}(w-i)
       \ind{q_i(x+\Delta)\ne q_i(y+\Delta)}.
\]
Since $2^i$ divides $2^{w-1}$, the probability of the displayed
indicator is $\min\{1,|x-y|/2^i\}$. Taking expectations gives
$2\sum_i(w-i)|x-y|=w(w+1)|x-y|$.
\end{proof}

\begin{proof}[Proof of Corollary~\ref{cor:l1}]
For $U=1$ every answer is zero. Otherwise draw the shift $\Delta$ of
Lemma~\ref{lem:l1-stability} before pattern preprocessing. For each
$\ell$, map every pattern and text symbol $x$ to $f_\ell(x+\Delta)$
and run the given Hamming matcher at accuracy $\eta$ on the mapped
stream. At each alignment $s$, estimate the mapped distance $H_\ell(s)$
by the median $\widehat H_\ell(s)$ of $O(\log J)$ independent copies.
Conditional on the shift, the mapped stream is fixed, so the median
fails with probability at most $1/(6J)$ for each fixed $s$.
The mapped windows all end at the same arrival. Form the signed estimate
$\widehat D_s=\sum_\ell c_\ell\widehat H_\ell(s)$ there.

By Lemma~\ref{lem:l1-stability},
$D_s=\sum_\ell c_\ell H_\ell(s)$ and the mass
$M_s=\sum_\ell|c_\ell|H_\ell(s)$ satisfies
$\E M_s\le J D_s$.
For fixed $s$ with $D_s>0$, Markov's inequality gives
$M_s\le6J D_s$ except with probability $1/6$. A union bound over the
$J$ medians gives failure probability at most $1/6$ for any shift.
When both events hold,
\[
 |\widehat D_s-D_s|\le\eta M_s\le\eps D_s/2.
\]
For $D_s=0$, every mapped distance and every successful Hamming
estimate is exactly zero. No cancellation assumption is needed.

For word arithmetic, clamp Hamming estimates to $[0,m]$ and round
them down to a common dyadic grid of mesh at most
$\eps/(2\sum_\ell|c_\ell|)$. The signed sum then has rounding
error at most $\eps/2$ and uses $O(\log(mU/\eps))=O(\log(mn))$ bits.
Zero is preserved, and every positive $D_s$ is an integer, so this
error fits the remaining budget.

Run $R=O(\log n)$ independent repetitions, each with its own shift, and
report their median at each alignment. Each repetition fails at a fixed
alignment with probability at most $1/3$, so median amplification and a
union bound give simultaneous success with probability at least $1-n^{-c}$.
There are $O(RJ\log J)$ Hamming instances, and all their retained states
count toward space. Each map takes constant word time. Computing each
median by linear-time selection, the reduction uses $O(RJ\log J)$
additional words and $O(RJ\log J)$ additional worst-case time per arrival.
\end{proof}

\end{document}